\documentstyle[amsfonts,amssymb,aps,multicol]{revtex}
\input{epsf}

\def\lsim{\mathrel{\raise.3ex\hbox{$<$\kern-.75em\lower1ex\hbox{$\sim$}}}}
\def\gsim{\mathrel{\raise.3ex\hbox{$>$\kern-.75em\lower1ex\hbox{$\sim$}}}}

\title{Confinement-Higgs transition in a disordered gauge theory \\
and the accuracy threshold for quantum memory\thanks{CALT-68-2374}}
\author{Chenyang Wang\thanks{chenyang@stanford.edu)}, Jim Harrington\thanks{\tt jimh@caltech.edu} and John Preskill\thanks{\tt
preskill@theory.caltech.edu}}

\address{Institute for Quantum Information, California Institute of Technology, 
Pasadena, CA 91125, USA}

\begin{document}

\maketitle

\begin{abstract}
We study the $\pm J$ random-plaquette $Z_2$ gauge model (RPGM) in three spatial dimensions, a three-dimensional analog of the two-dimensional $\pm J$ random-bond Ising model (RBIM). The model is a pure $Z_2$ gauge theory in which randomly chosen plaquettes (occuring with concentration $p$) have couplings with the ``wrong sign'' so that magnetic flux is energetically favored on these plaquettes. Excitations of the model are one-dimensional ``flux tubes'' that terminate at ``magnetic monopoles'' located inside lattice cubes that contain an odd number of wrong-sign plaquettes. Electric confinement can be driven by thermal fluctuations of the flux tubes, by the quenched background of magnetic monopoles, or by a combination of the two. Like the RBIM, the RPGM has enhanced symmetry along a ``Nishimori line'' in the $p$-$T$ plane (where $T$ is the temperature). The critical concentration $p_c$ of wrong-sign plaquettes at the confinement-Higgs phase transition along the Nishimori line can be identified with the accuracy threshold for robust storage of quantum information using topological error-correcting codes: if qubit phase errors, qubit bit-flip errors, and errors in the measurement of local check operators all occur at rates below $p_c$, then encoded quantum information can be protected perfectly from damage in the limit of a large code block. Through Monte-Carlo simulations, we measure $p_{c0}$, the critical concentration along the $T=0$ axis (a lower bound on $p_c$), finding $p_{c0}=.0293\pm .0002$. We also measure the critical concentration of antiferromagnetic bonds in the two-dimensional RBIM on the $T=0$ axis, finding $p_{c0}=.1031\pm .0001$. Our value of $p_{c0}$ is incompatible with the value of $p_c=.1093\pm .0002$ found in earlier numerical studies of the RBIM, in disagreement with the conjecture that the phase boundary of the RBIM is vertical (parallel to the $T$ axis) below the Nishimori line. The model can be generalized to a rank-$r$ antisymmetric tensor field in $d$ dimensions, in the presence of quenched disorder.
\end{abstract}

\begin{multicols}{2}[]
\section{Introduction}
Spin systems with quenched randomness have been extensively studied, leading to valuable insights that apply to (for example) spin glass materials, quantum Hall systems, associative memory, error-correcting codes, and combinatorial optimization problems \cite{parisi_book,young_book,nishi_book}. Gauge systems with quenched randomness, which have received comparatively little attention, will be studied in this paper. 

The gauge models we consider are intrinsically interesting because they provide another class of simple systems with disorder-driven phase transitions. But our investigation of these models has a more specific motivation connected to the theory of quantum error correction. 

In practice, coherent quantum states rapidly decohere due to uncontrollable interactions with the environment. But in principle, if the quantum information is cleverly encoded \cite{shor_9,steane_7}, it can be stabilized and preserved using fault-tolerant recovery protocols \cite{shor_ft}.  Kitaev \cite{kitaev1,kitaev2} proposed a particularly promising class of quantum error-correcting codes ({\em surface codes}) in which the quantum processing required for error recovery involves only {\em local} interactions among qubits arranged in a two-dimensional block, and the protected information is associated with global topological properties of the quantum state of the block. If the error rate is small, then the topological properties of the code block are well protected, and error recovery succeeds with a probability that rapidly approaches one in the limit of a large code block. But if the error rate is above a critical value, the {\em accuracy threshold}, then quantum error correction is ineffective.

In \cite{topo_memory}, a precise connection was established between the accuracy threshold achievable with surface codes and the confinement-Higgs transition in a three-dimensional $Z_2$ lattice gauge model with quenched randomness. The model has two parameters: the temperature $T$ and the concentration $p$ of ``wrong-sign'' plaquettes. On wrong-sign plaquettes (which are analogous to antiferromagnetic bonds in a spin system) it is energetically favorable for the $Z_2$ magnetic flux to be nontrivial. In the mapping between quantum error recovery and the gauge model, the quenched fluctuations correspond to the actual errors introduced by the environment; these impose sites of frustration,  {\em magnetic monopoles}, corresponding to an ``error syndrome'' that can be measured by executing a suitable quantum circuit.  Thermally fluctuating magnetic flux tubes, which terminate at magnetic monopoles, correspond to the ensemble of possible error patterns that could generate a particular error syndrome. (The temperature $T$ is tied to the strength $p$ of the quenched fluctuations through a {\em Nishimori relation} \cite{nish_line}.) When the disorder is weak and the temperature low (corresponding to a small error rate), the system is in a magnetically ordered Higgs phase. In the surface code, magnetic order means that all likely error patterns that might have produced the observed error syndrome are topologically equivalent, so that the topologically encoded information resists damage. But at a critical value $p_c$ of the disorder strength (and a temperature determined by Nishimori's relation), magnetic flux tubes condense and the system enters the magnetically disordered confinement phase. In the surface code, magnetic disorder means that the error syndrome cannot point to likely error patterns belonging to a unique topological class; therefore topologically encoded information is vulnerable to damage.

Although the code block is two dimensional, the gauge model is three dimensional because one dimension represents {\em time}. Time enters the analysis of recovery because measurements of the error syndrome might themselves be faulty; therefore measurements must be repeated on many successive time slices if they are to provide reliable information about the errors that have afflicted the code block. If qubit phase errors, qubit bit-flip errors, and errors in the measurement of local check operators all occur at rates below $p_c$, then encoded quantum information can be protected perfectly from damage in the limit of a large code block. As we consider more and more reliable measurements of the syndrome, the corresponding three-dimensional gauge model becomes more and more anisotropic, reducing in the limit of perfect measurements to the two-dimensional random-bond Ising model.

The numerical value $p_c$ of the accuracy threshold is of considerable interest, since it characterizes how reliably quantum hardware must perform in order for a quantum memory to be robust. In the three-dimensional $Z_2$ gauge model, $p_c$ is the value of the wrong-sign pla\-quette concentration where the confinement-Higgs boundary crosses the Nishimori line in the $p$-$T$ plane. A lower bound on $p_c$ is provided by the critical concentration $p_{c0}$ on the $T=0$ axis. In \cite{topo_memory}, an analytic argument established that $p_{c0}\ge .0114$. In this paper we report on a numerical calculation that finds $p_{c0}=.0293\pm .0002$. 

In the case where the error syndrome can be measured flawlessly, the critical error rate is given by the critical antiferromagnetic bond concentration on the Nishimori line of the two-dimensional random-bond Ising model (RBIM). Numerical calculations performed earlier by other authors \cite{honecker,chalker} have established $p_c=.1093\pm .0002$. According to a conjecture of Nishimori \cite{nish_vert} and Kitatani \cite{kitatani}, this value of $p_c$ should agree with the critical bond concentration $p_{c0}$ of the 2D RBIM on the $T=0$ axis. The same reasoning that motivates this conjecture for the  RBIM indicates that $p_c=p_{c0}$ for the 3D random-plaquette gauge model (RPGM) as well. However, we have calculated $p_{c0}$ in the 2D RBIM numerically, finding $p_{c0}=.1031\pm .0001$. Our value of $p_{c0}$ agrees with an earlier numerical calculation by Kawashima and Rieger \cite{kawashima}, but disagrees with the conjecture that $p_{c}=p_{c0}$.

In Sec. II we describe in more detail the properties of the 2D RBIM and the 3D RPGM, emphasizing the importance of the Nishimori line and the inferences that can be made about the behavior of order parameters on this line. Section III reviews the connection between the models and error recovery using surface codes. Our numerical results for $p_{c0}$ and for the critical exponent $\nu_0$ at the $T=0$ critical point are presented in Sec. IV. Section V summarizes our conclusions.

\section{Models}

\subsection{Random-bond Ising model}
The two-dimensional $\pm J$ random-bond Ising model (RBIM) has a much studied multicritical point at which both the temperature and the strength of quenched disorder are nonzero. This model is an Ising spin system on a square lattice, with a variable $S_i=\pm 1$ residing at each lattice site $i$. Its Hamiltonian is 
\begin{equation}
\label{Ising_Hamiltonian}
H= -J \sum_{\langle ij\rangle}\tau_{ij}S_i S_j~,
\end{equation}
where $J$ is the strength of the coupling between neighboring spins, and $\tau_{ij}=\pm 1$ is a quenched random variable. (That is, $\tau_{ij}$ depends on what {\em sample} of the system is selected from a certain ensemble, but is not subject to thermal fluctuations.) The $\tau_{ij}$'s are independently and identically distributed, with the antiferromagnetic choice $\tau_{ij}=-1$ (favoring that neighboring spins antialign) occuring with probability $p$, and the ferromagnetic choice $\tau_{ij}$=+1 (favoring that neighboring spins align) occuring with probability $1-p$. We refer to $p$ as the concentration of antiferromagnetic bonds, or simply the bond concentration. 

The free energy $F$ of the model at inverse temperature $\beta$, averaged over samples, is
\begin{equation}
\label{free_energy}
[\beta F(K,\tau)]_{K_p}=-\sum_{\tau}P(K_p,\tau)\ln Z(K,\tau)
\end{equation}
where
\begin{equation}
\label{partition_function}
Z(K,\tau)=\sum_{S}\exp\left(K\sum_{\langle ij\rangle}\tau_{ij}S_i S_j\right)
\end{equation}
is the partition function for sample $\tau$ (with $K=\beta J$), and
\begin{equation}
P(K_p,\tau)= (2\cosh K_p)^{-N_B}\times \exp\left(K_p \sum_{\langle ij\rangle}\tau_{ij}\right)
\end{equation}
is the probability of the sample $\tau$; here 
\begin{equation}
{p\over 1-p}=e^{-2K_p}
\end{equation}
and $N_B$ is the number of bonds.

The partition function $Z(K,\tau)$ is invariant under the change of variable
\begin{equation}
\label{var_change}
S_i\to \sigma_i S_i~,\quad \tau_{ij}\to \sigma_i\sigma_j\tau_{ij}~,
\end{equation}
where $\sigma_i=\pm 1$. Thus $\tau$ itself has no invariant meaning --- samples $\tau$ and $\tau'$ that differ by the change of variable have equivalent physics. The only invariant property of $\tau$ that cannot be modified by such a change of variable is the {\em distribution of frustration} that $\tau$ determines. If an odd number of the bonds contained in a specified plaquette have $\tau=-1$ then that plaquette is frustrated --- an {\em Ising vortex} resides at the plaquette. For purposes of visualization, we sometimes will find it convenient to define the spin model on the dual lattice so that the spins reside on plaquettes and the Ising vortices reside on sites. Then excited bonds with $\tau_{ij}S_iS_j = -1$ form one-dimensional chains that terminate at the frustrated sites. 

Changes of variable define an equivalence relation on the set of $2^{N_B}$ $\tau$ configurations: there are the $2^{N_S}$ elements of each equivalence class (the number of changes of variable, where $N_S$ is the number of sites) and there are $2^{N_S}$ classes (the number of configurations for the Ising vortices --- note that $N_B=2N_S$ for a square lattice on the 2-torus, and that the number of plaquettes is $N_P=N_S$). Denote a distribution of Ising vortices, or equivalently an equivalence class of $\tau$'s, by $\eta$. The probability $P(K_p,\eta)$ of $\eta$ is found by summing $P(K_p,\tau)$ over all the representatives of the class; hence
\begin{eqnarray}
\label{class_prob}
&&(2\cosh K_p)^{N_B}P(K_p,\eta)=  (2\cosh K_p)^{N_B}\sum_{\tau\in\eta}P(K_p,\tau)\nonumber\\
&&=\sum_\sigma\exp\left(K_p \sum_{\langle ij\rangle}\tau_{ij}\sigma_i\sigma_j\right)=Z(K_p,\eta)~.
\end{eqnarray}
Apart from a normalization factor, the probability of a specified distribution of frustration is given by the partition function of the model, but with $K=\beta J$ replaced by $K_p$.

In this model, we can define an order parameter that distinguishes the ferromagnetic and paramagnetic phases. Let
\begin{equation}
m^2(K,K_p)=\lim_{|i-j|\to \infty}\left[\left\langle S_i S_j\right\rangle_K\right]_{K_p}~,
\end{equation}
where $\langle\cdot\rangle_K$ denotes the average over thermal fluctuations, $[\cdot ]_{K_p}$ denotes the average over samples, and $|i-j|$ denotes the distance between site 
$i$ and site $j$; then in the ferromagnetic phase $m^2 >0$ and in the paramagnetic phase $m^2=0$. But the two-point correlation function $\langle S_i S_j\rangle_K$ is not invariant under the change of variable eq.~(\ref{var_change}), so how should $m^2$ be interpreted?

\begin{figure}
\begin{center}
\leavevmode
\epsfxsize=3in
\epsfbox{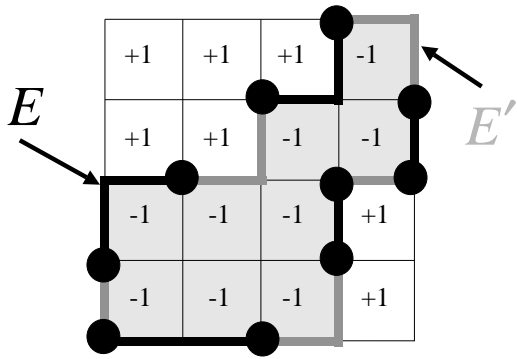}
\end{center}
\caption{The  chain $E$ of antiferromagnetic bonds (darkly shaded) and the  chain $E'$ of excited bonds (lightly shaded), in the two-dimensional random-bond Ising model. Ising spins taking values in $\{\pm 1\}$ reside on plaquettes; Ising vortices (boundary points of $E$) are located on the sites marked by filled circles. The bonds of $E'$ comprise a one-dimensional defect that connects the vortices. The cycle $D=E+E'$ encloses a domain of spins with the value $-1$. }
\label{fig:rbim}
\end{figure}

Following \cite{topo_memory}, denote by $E$ the set of bonds that are antiferromagnetic ($\tau_{ij}=-1$), denote by $E'$ the set of excited bonds with $\tau_{ij}S_i S_j=-1$, and denote by  $D$ the set of bonds with $S_i S_j=-1$ (those such that the neighboring spins antialign) --- see Fig.~\ref{fig:rbim}. Then $D=E+E'$ is the disjoint union of $E$ and $E'$ (containing bonds in $E$ or $E'$ but not both). Furthermore, $D$ contains an even number of the bonds that meet at any given site; that is, $D$ is a {\em cycle}, a chain of bonds that has no boundary points. The quantity $S_i S_j$
just measures whether a line connecting $i$ and $j$ crosses $D$ an even number ($S_i S_j=1$) or an odd number ($S_i S_j=-1$) of times. 

Now $D$ consists of disjoint ``domain walls'' that form closed loops. If loops that are arbitrarily large appear with appreciable weight in the thermal ensemble, then the two-point function $\langle S_i S_j\rangle_K$ decays like $\exp(-|i-j|/\xi)$ --- fluctuations far from the sites $i$ and $j$ contribute to the correlation function. Thus the spins are disordered and $m^2=0$. But if large loops occur only with negligible probability, then only fluctuations localized near $i$ and $j$ contribute significantly; the spin correlation persists at large distances and $m^2>0$. Thus, the order parameter probes whether the chain $E'$ of excited bonds can wander far from the chain $E$ of ferromagnetic bonds; that is, whether $D=E+E'$ contains arbitrarily large connected closed loops, for typical thermal fluctuations and typical samples.

Nishimori \cite{nish_line} observed that the model has enhanced symmetry properties along a line in the $p$-$T$ plane (the {\em Nishimori line}) defined by $K=K_p$ or $\exp(-2\beta J)= p/(1-p)$. In this case, the antiferromagnetic bond chain $E$ and the excited bond chain $E'$ are generated by sampling the same probability distribution, subject to the constraint that both chains have the same boundary points. This feature is preserved by renormalization group transformations, so that renormalization group flow preserves Nishimori's line \cite{ledoussal}. The {\em Nishimori point} $(p_c,T_c)$ where the Nishimori line crosses the ferromagnetic-paramagnetic phase boundary, is a renormalization group fixed point, the model's multicritical point. 

When the temperature $T$ is above the Nishimori line, excited bonds have a higher concentration than antiferromagnetic bonds, so we may say that thermal fluctuations play a more important role than quenched randomness in disordering the spins. When $T$ is below the Nishimori line, antiferromagnetic bonds are more common than excited bonds, and the quenched randomness dominates over thermal fluctuations. Right on the Nishimori line, the effects of thermal fluctuations and quenched randomness are in balance \cite{nish_2002}.

By invoking the change of variable eq.~(\ref{var_change}), various properties of the model on the Nishimori line can be derived \cite{nishi_book,nish_line}. For example, the internal energy density (or ``average bond'') can be computed analytically, 
\begin{equation}
[\tau_{ij}\langle S_i S_j\rangle_{K_p}]_{K_p}= 1-2p~,
\end{equation}
where $i$ and $j$ are neighboring sites; averaged over thermal fluctuations and samples, the concentration of excited bonds is $p$ as one would expect (and the internal energy has no singularity at the Nishimori point).  Furthermore, after averaging over disorder, the $(2m-1)$st power of the $k$-spin correlator has the same value as the $(2m)$th power, for any positive integer $m$:
\begin{equation}
\left[\langle S_{i_1} S_{i_2}\cdots S_{i_k}\rangle_{K_p}^{2m-1}\right]_{K_p}
= \left[\langle S_{i_1} S_{i_2}\cdots S_{i_k}\rangle_{K_p}^{2m}\right]_{K_p}~.
\end{equation}
It follows in particular that the spin-glass order parameter
\begin{equation}
q^2(K_p,K_p)\equiv \lim_{|i-j|\to \infty}\left[\langle S_i S_j\rangle_{K_p}^2\right]_{K_p}
\end{equation}
coincides with the ferromagnetic order parameter $m^2(K_p,K_p)$ along the Nishimori line, reflecting the property that thermal fluctuations and quenched randomness have equal strength on this line.

\begin{figure}
\begin{center}
\leavevmode
\epsfxsize=3.25in
\epsfbox{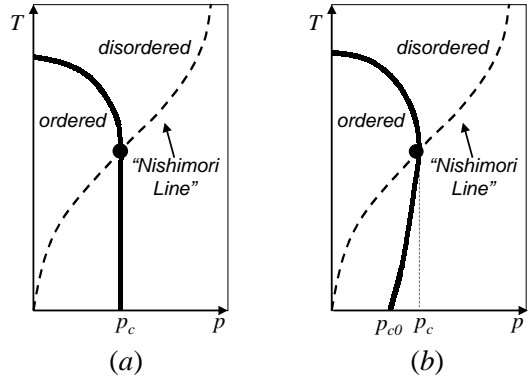}
\end{center}
\caption{The phase diagram of the random-bond Ising model (shown schematically), with the temperature $T$ on the vertical axis and the concentration $p$ of antiferromagnetic bonds on the horizontal axis. The solid line is the boundary between the ferromagnetic (ordered) phase and the paramagnetic (disordered) phase. The dotted line is the Nishimori line $e^{-2\beta J}= p/(1-p)$, which crosses the phase boundary at the Nishimori point (the heavy black dot). It has been conjectured, but not proven, that the phase boundary from the Nishimori point to the $p$-axis is {\em vertical}, as in ($a$). The numerics reported in Sec.~IV favor the reentrant phase diagram shown in ($b$). The deviation of the critical bond concentration $p_c$ on the Nishimori line from the critical bond concentration $p_{c0}$ on the $T=0$ axis has been exaggerated in ($b$) for clarity.}
\label{fig:nishi}
\end{figure}

Comparing eq.~(\ref{free_energy}) and (\ref{class_prob}), we see that for $K=K_p$ the free energy of the model coincides with the {\em Shannon entropy} of the distribution of vortices, apart from a nonsingular additive term:
\begin{eqnarray}
&&\left[\beta F(K_p,\tau)\right]_{K_p} \nonumber\\
&&= -\sum_\eta P(K_p,\eta)\ln P(K_p,\eta) - N_B\ln\left(2\cosh K_p\right)~.
\end{eqnarray}
Since the free energy is singular at the Nishimori point $(p_c,T_c)$, it follows that the Shannon entropy of frustration (which does not depend on the temperature) is singular at $p=p_c$ \cite{nish_vert}. This property led Nishimori to suggest that the boundary between the ferromagnetic and paramagnetic phases occurs at $p=p_c$ at sufficiently low temperature, and thus that the phase boundary is vertical in the $p$-$T$ plane below the Nishimori point, as in Fig.~\ref{fig:nishi}a. Later, Kitatani \cite{kitatani} arrived at the same conclusion by a different route, showing that the verticality of the phase boundary follows from an ``appropriate condition.'' These arguments, while suggestive, do not seem compelling to us. There is no known rigorous justification for Kitatani's condition, and no rigorous reason why the ferro-para boundary must coincide with the singularity in the entropy of frustration, even at low temperature. Hence we regard the issue of the verticality of the phase boundary as still unsettled. Nishimori did argue convincingly that the phase boundary cannot extend to any value of $p$ greater than $p_c$ \cite{nish_line}, and Le Doussal and Harris argued that the {\em tangent} to the phase boundary is vertical at the Nishimori point \cite{ledoussal}, but these results leave open the possibility of a ``reentrant'' boundary that slopes back toward the $T$ axis below the Nishimori point, as in Fig.~\ref{fig:nishi}b.

The RBIM can also be defined in $d$ dimensions. Much of the above discussion still applies, with minor modifications. Consider, for example, $d=3$. On the dual lattice, spins reside on lattice cubes and the bonds become plaquettes shared by two neighboring cubes. The set of antiferromagnetic bonds $E$ is dual to a two-dimensional surface, and its boundary $\partial E$ consists of  one-dimensional loops --- the Ising {\em strings} where the spins are frustrated. The set of excited bonds $E'$ is dual to another two-dimensional surface that is also bounded by the Ising strings: $\partial E'=\partial E$. The spins are disordered if the two-cycle $D=E+E'$ contains arbitrarily large closed connected surfaces for typical thermal fluctuations and typical samples. Similarly, in $d$ dimensions, frustration is localized on closed surfaces of dimension $d-2$, and the thermally fluctuating defects are dimension-$(d-1)$ surfaces that terminate on the locus of frustration. For any $d$, the model has enhanced symmetry along the Nishimori line $K=K_p$, where antiferromagnetic bonds and excited bonds are drawn from the same probability distribution.

In the absence of quenched disorder, the two-dimensional Ising model is mapped to itself by a duality relation that can be used to infer properties of the critical theory. When quenched disorder is introduced, however, the two-dimensional random bond Ising model is mapped under duality to a model with Boltzmann weights that are not positive definite \cite{nishi_dual}, so that it is not easy to draw any firm conclusions.

\subsection{Random-plaquette gauge model}

In the $d$-dimensional RBIM, excitations have codimension 1 and terminate on a closed surface of codimension 2. The $Z_2$ random-plaquette gauge model (RPGM) is defined in an entirely analogous manner, except that the excitations are codimension-2 objects (``magnetic flux tubes'') that terminate on codimension-3 objects (``magnetic monopoles'').

More concretely, the variables of the model are $U_\ell=\pm 1$ residing on each link $\ell$ of the lattice, and the Hamiltonian is 
\begin{equation}
H= -J \sum_{P}\tau_P U_P~,
\end{equation}
where $J$ is the coupling strength, 
\begin{equation}
U_P = \prod_{\ell\in P}U_\ell
\end{equation}
is the $Z_2$-valued ``magnetic flux'' through the plaquette $P$, and $\tau_P=\pm 1$ is a quenched random variable.  The $\tau_P$'s are independently and identically distributed, with the ``wrong-sign''  choice $\tau_P=-1$ (favoring nontrivial flux) occuring with probability $p$, and the ``right-sign'' choice $\tau_P$=+1 (favoring trivial flux) occuring with probability $1-p$. We refer to $p$ as the concentration of wrong-sign plaquettes, or simply the plaquette concentration. 

The free energy $F$ of the model at inverse temperature $\beta$, averaged over samples, is
\begin{equation}
\label{free_energy_gauge}
[\beta F(K,\tau)]_{K_p}=-\sum_{\tau}P(K_p,\tau)\ln Z(K,\tau)
\end{equation}
where
\begin{equation}
Z(K,\tau)=\sum_{U}\exp\left(K\sum_{P}\tau_P U_P\right)
\end{equation}
is the partition function for sample $\tau$ (with $K=\beta J$), and
\begin{equation}
P(K_p,\tau)= (2\cosh K_p)^{-N_P}\times \exp\left(K_p \sum_{P}\tau_P \right)
\end{equation}
is the probability of the sample $\tau$; here 
\begin{equation}
{p\over 1-p}=e^{-2K_p}
\end{equation}
and $N_P$ is the number of plaquettes.

The partition function $Z(K,\tau)$ is invariant under the change of variable
\begin{equation}
\label{var_change_gauge}
U_\ell\to \sigma_\ell U_\ell~,\quad \tau_{P}\to \sigma_P\tau_{P}~,
\end{equation}
where $\sigma_\ell=\pm 1$ and $\sigma_P=\prod_{\ell\in P}\sigma_\ell$. While $\tau$ itself has no invariant meaning, $\tau$ determines a distribution of frustration that cannot be altered by a change of variable. If an odd number of the plaquettes contained in a specified cube have $\tau=-1$ then that cube is frustrated --- a $Z_2$ {\em magnetic monopole} resides in the cube. For purposes of visualization, we will sometimes find it convenient to define the gauge model on the dual lattice so that the gauge variables $U_\ell$ reside on plaquettes, the magnetic flux on bonds, and the magnetic monopoles on sites. Then excited bonds with $\tau_{P}U_P = -1$ form one-dimensional strings that terminate at monopoles. 

We can define an order parameter that distinguishes the Higgs (magnetically ordered) phase and the confinement (magnetically disordered) phase. Consider the Wilson loop operator associated with a closed loop $C$ (on the original lattice, not the dual lattice):
\begin{equation}
W(C)=\prod_{\ell\in C}U_\ell~.
\end{equation}
and consider the behavior of the expectation value of $W(C)$, averaged over thermal fluctuations and over samples. In the Higgs phase, for a large loop $C$ the Wilson loop operator decays exponentially with the perimeter of the loop,
\begin{equation}
\left[\langle W(C)\rangle_{K}\right]_{K_p}\sim \exp\left[-\mu\cdot {\rm Perimeter}(C)\right]~,
\end{equation}
while in the confinement phase it decays exponentially with the area of the minimal surface bounded by $C$,
\begin{equation}
\left[\langle W(C)\rangle_{K}\right]_{K_p}\sim \exp\left[-\kappa\cdot {\rm Area}(C)\right]~.
\end{equation}
The interpretation is that on the dual lattice the wrong-sign plaquettes correspond to a one-chain $E$ bounded by magnetic monopoles, and the excited plaquettes correspond to another one-chain $E'$ with the same boundary; hence $D=E+E'$ is a cycle, a sum of disjoint closed ``flux tubes.'' If arbitrarily large loops of flux appear with appreciable weight in the thermal ensemble for typical samples, then magnetic fluctuations spanning the entire surface bounded by $C$ contribute to the expectation value of $W(C)$, and the area-law decay results. If large flux tubes are suppressed, then only the fluctuations localized near the loop are important, and the perimeter-law decay applies. Thus, the Wilson-loop order parameter probes whether the chain $E'$ of excited plaquettes can wander far from the chain $E$ of wrong-sign plaquettes; that is, whether $D=E+E'$ contains arbitrarily large connected closed loops.

The one-chain $E$ bounded by the magnetic monopoles is analogous to a $Z_2$-valued {\em Dirac string} --- the change of variable eq.~(\ref{var_change_gauge}) deforms the strings while leaving invariant the boundary of $E$ (the locations of the monopoles). One should notice that these strings are {\em not} invisible to our Wilson loop operator; that is $W(C)$ is not invariant under the change of variable. It is possible to modify $W(C)$ to obtain an invariant object \cite{alford}, but that would not be appropriate if the order parameter is supposed to probe the extent to which the thermally fluctuating defects (the excited plaquettes) depart from the quenched disorder (the Dirac strings).

Like the RBIM, the RPGM has enhanced symmetry on the Nishimori line $K=K_p$, and the change of variable eq.~(\ref{var_change_gauge}) may be invoked to derive properties of the model on this line. The Nishimori line is preserved by renormalization group flow, and crosses the confinement-Higgs boundary at a multicritical point $(p_c,T_c)$. The internal energy (or average plaquette) can be computed on this line,
\begin{equation}
\left[\tau_P\langle U_P\rangle_{K_p}\right]_{K_p}= 1-2p~
\end{equation}
(excited plaquettes have concentration $p$) and for each positive integer $m$, the $(2m-1)$'st power of W(C) and the $2m$'th power are equal when averaged over samples,
\begin{equation}
\label{wilson_ident}
\left[\langle W(C)\rangle_{K_p}^{2m-1}\right]_{K_p}=\left[\langle W(C)\rangle_{K_p}^{2m}\right]_{K_p}~.
\end{equation}
Furthermore, the free energy on the Nishimori line, apart from a nonsingular additive term, is equal to the Shannon entropy of the distribution of magnetic monopoles, so that the latter is singular at $p=p_c$. 

In principle, the RPGM could have what might be called a ``gauge glass'' phase. In this phase, the Wilson loop, averaged over thermal and quenched fluctuations, has area-law behavior,
\begin{equation}
\left[\langle W(C)\rangle_{K}\right]_{K_p}\sim \exp\left[-\kappa\cdot {\rm Area}(C)\right]~,
\end{equation}
but the {\em square} of its thermal expectation value, averaged over quenched fluctuations, has perimeter-law behavior:
\begin{equation}
\left[\langle W(C)\rangle_{K}^2\right]_{K_p}\sim \exp\left[-\mu\cdot {\rm Perimeter}(C)\right]~.
\end{equation}
This means that thermal fluctuations do not induce magnetic disorder for each typical sample, but that the magnetic fluctuations are large when we compare one sample to another.  However, the identity eq.~(\ref{wilson_ident}) shows that, along the Nishimori line $K=K_p$, there can be no gauge glass phase. Since $\langle W(C)\rangle$ and $\langle W(C)\rangle^2$ have the same average over samples, both order parameters cross from perimeter to area law at the same point on the Nishimori line. (Nishimori \cite{nish_line} used the analogous argument to show that there is no spin glass behavior in the RBIM along the Nishimori line.)

Another useful identity that can be derived using the change of variable is 
\begin{equation}
\left[\langle W(C)\rangle_{K}\right]_{K_p}=\left[\langle W(C)\rangle_{K} \langle W(C)\rangle_{K_p}\right]_{K_p}~.
\end{equation}
Since $-1\le W(C) \le 1$, it follows that
\begin{equation}
\label{wilson_abs}
\left|~\left[\langle W(C)\rangle_{K}\right]_{K_p}~\right|\le\left[~\left|\langle W(C)\rangle_{K_p}\right|~\right]_{K_p}~.
\end{equation}
From this inequality, we may infer that if the point on the Nishimori line with concentration $p$ is in the confinement phase, then the point $(p,T)$ is in the confinement phase for any temperature $T$. (Again, the reasoning is exactly analogous to Nishimori's argument for the RBIM \cite{nish_line}.) Since there is no gauge-glass behavior on the Nishimori line, if a point on the Nishimori line is in the confinement phase, then $\langle W(C)\rangle_{K_p}$ already exhibits area-law decay before averaging over samples. Therefore the right-hand side of eq.~(\ref{wilson_abs}) shows area-law decay and so must the left-hand side. We conclude that, as for the RBIM, the phase boundary of the RPGM below the Nishimori line must either be vertical (parallel to the $T$ axis as in Fig.~\ref{fig:nishi}a) or reentrant (tipping back toward the $T$ axis as $T$ decreases as in Fig.~\ref{fig:nishi}b). 

\subsection{Further generalizations}
In $d$ dimensions, the magnetic order parameter of the RBIM explores whether a thermally excited chain $E'$ of codimension 1 (domain walls) deviates far from a quenched codimension-1 chain $E$ (antiferromagnetic bonds), where both $E$ and $E'$ have the same codimension-2 boundary (the Ising vortices). Similarly, the RPGM can be defined in $d$ dimensions, and its Wilson-loop order parameter probes whether a thermally excited chain $E'$ of codimension 2 (flux tubes) deviates far from a quenched codimension-2 chain $E$ (Dirac strings), where both $E$ and $E'$ have the same codimension-3 boundary (the magnetic monopoles).

This concept admits further generalizations. In $d$-dimensions, we may consider the lattice theory of a ``rank-$r$ antisymmetric tensor field'' with quenched disorder. Then variables reside on the $r$-cells of the lattice, and the Hamiltonian is expressed in terms of a field strength defined on $(r+1)$-cells. The sign of the coupling is determined by a random variable $\tau$ taking values $\pm 1$ on $(r+1)$-cells; cells with the ``wrong sign'' have concentration $p$. On the dual lattice, $\tau$ corresponds to a codimension-$(r+1)$ chain $E$, and the excited cells to a codimension-$(r+1)$ chain $E'$, where $E$ and $E'$ are bounded by the same codimension-$(r+2)$ chain of frustration. An operator analogous to the Wilson loop can be defined that detects the flux through the dimension-($r+1$) ``surface'' bounded by a dimension-$r$ ``loop'' $C$; this operator serves as the order parameter for an order-disorder transition. The order parameter probes whether the thermally fluctuating codimension-$(r+1)$ chain $E'$ deviates far from the quenched codimension-$(r+1)$ chain $E$.

For any $d$ and $r$, the model has enhanced symmetry on the Nishimori line, where $K=K_p$. Properties of the model on this line can be derived, analogous to those discussed above for the RBIM and the RPGM.

\section{Accuracy threshold for quantum memory}

How the RBIM and RPGM relate to the performance of topological quantum memory was extensively discussed in \cite{topo_memory}. Here we will just briefly reprise the main ideas.

\subsection{Toric codes}
Quantum information can be protected from decoherence and other possible sources of error using quantum error-correcting codes \cite{shor_9,steane_7} and fault-tolerant error recovery protocols \cite{shor_ft}. Topological codes (or {\em surface codes}) are designed so that the quantum processing needed to control errors has especially nice locality properties \cite{kitaev1,kitaev2}. 
 
\begin{figure}
\begin{center}
\leavevmode
\epsfxsize=3in
\epsfbox{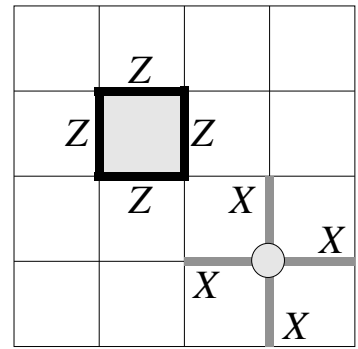}
\end{center}
\caption{The check operators of the toric code. Each plaquette operator is a tensor product of $Z$'s acting on the four links contained in the plaquette. Each site operator is a tensor product of $X$'s acting on the four links that meet at the site.}
\label{fig:checks}
\end{figure}

Specifically, consider a system of $2L^2$ qubits (a qubit is a two-level quantum system), with each qubit residing at a link of an $L\times L$ square lattice drawn on a two-dimensional {\em torus}. (Other examples of surface codes, including codes defined on planar surfaces, are discussed in \cite{topo_memory}.) This system can encode two qubits of quantum information that are well protected from noise if the error rate is low enough. The two-qubit code space, where the protected information resides, can be characterized as a simultaneous eigenspace with eigenvalue one of a set of check operators (or ``stabilizer generators''); check operators are associated with each site and with each elementary cell (or ``plaquette'') of the lattice, as shown in Fig.~\ref{fig:checks}. We use the notation
\begin{eqnarray}
& &I= \pmatrix{1 & 0\cr 0 & 1\cr}~,~\quad X=\pmatrix{0 & 1\cr 1 & 0\cr}~,\\
& &Y= \pmatrix{0 & -i\cr i & 0\cr}~, ~Z=\pmatrix{1 & 0\cr 0 & -1\cr}
\end{eqnarray}
for the $2\times 2$ identity and Pauli matrices. The check operator at site $i$ acts nontrivially on the four links that meet at the site; it is the tensor product 
\begin{equation}
X_i=\otimes_{\ell\ni s} X_\ell
\end{equation}
acting on those four qubits, times the identity acting on the remaining qubits. The check operator at plaquette $P$ acts nontrivially on the four links contained in the plaquette, as the tensor product 
\begin{equation}
Z_P=\otimes_{\ell\in P} Z_\ell~,
\end{equation}
times the identity on the remaining links. 

The check operators can be simultaneously diagonalized, and the toric code is the space in which each check operator acts trivially.  Because of the periodic boundary conditions on the torus, the product of all $L^2$ site operators or all $L^2$ plaquette operators is the identity --- each link operator occurs twice in the product, and $X^2=Z^2=I$. There are no further relations among these operators; therefore, there are $2\cdot (L^2-1)$ independent check operators constraining the $2L^2$ qubits in the code block, and hence two encoded qubits (the code subspace is four dimensional).

Since the check operators are spatially local, it is useful to think of a site or plaquette where the check operator has the eigenvalue $-1$ as the position of a localized excitation or ``defect.'' The code space contains states with no defects, which are analogous to vacuum states of a $Z_2$ gauge theory on the torus: $Z_P=1$ means that there is no $Z_2$ magnetic flux at plaquette $P$, and $X_i=1$ means that there is no $Z_2$ electric charge at site $i$. (This $Z_2$ gauge theory on the two-torus should not be confused with the three-dimensional $Z_2$ gauge theory, described in Sec.~\ref{sec:memory_rpgm}, that arises in the analysis of the efficacy of error correction!)

\begin{figure}
\begin{center}
\leavevmode
\epsfxsize=3in
\epsfbox{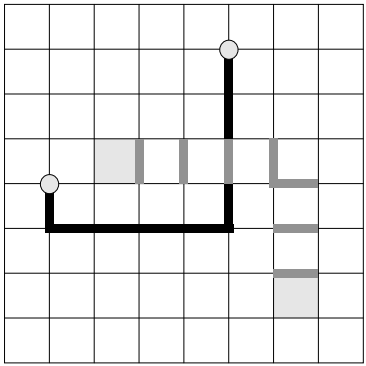}
\end{center}
\caption{Site defects and plaquette defects in the toric code. Applied to the code space, $Z$'s acting on a connected chain of links (darkly shaded) create site defects (electric charges) at the ends of the chain. Similarly, $X$'s applied to a connected chain of dual links (lightly shaded) create plaquette defects (magnetic fluxes) at the ends of the chain.}
\label{fig:defects}
\end{figure}

Consider applying to the vacuum state an operator that is a tensor product of Pauli matrices $\{Z_\ell\}$ acting on each of a set of links forming a connected chain $\{\ell\}$. This operator creates isolated site defects at the ends of the chain. Similarly, if we apply to the vacuum a tensor product of Pauli matrices $\{X_\ell\}$ acting on a connected chain of the dual lattice, isolated plaquette defects are created at the ends of the chain, as in Fig.~\ref{fig:defects}. A general ``Pauli operator'' (tensor product of Pauli matrices) can be expressed as tensor product of $X_\ell$'s and $I_\ell$'s times a tensor products of $Z_\ell$'s and $I_\ell$'s; this operator preserves the code space if and only if the links acted upon by $Z$'s comprise a {\em cycle} of the lattice (a chain with no boundary) and the links acted upon by $X$'s comprise a cycle of the dual lattice. 

\begin{figure}
\begin{center}
\leavevmode
\epsfxsize=3in
\epsfbox{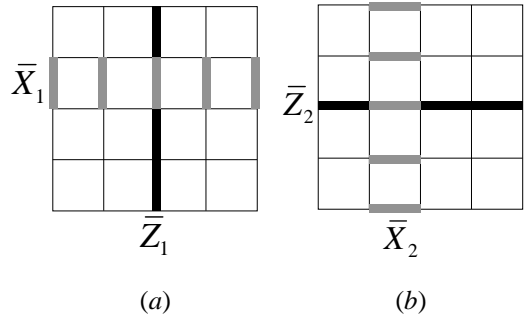}
\end{center}
\caption{Basis for the operators that act on the two encoded qubits of the toric code. ($a$) The encoded $\bar Z_1$ is a tensor product of $Z$'s acting on lattice links comprising a cycle of the torus, and the encoded $\bar X_1$ is a tensor product of $X$'s acting on dual links comprising the complementary cycle. ($b$) $\bar Z_2$ and $\bar X_2$ are defined similarly.}
\label{fig:encoded_ops}
\end{figure}

Cycles on the torus are of two types. A {\em homologically trivial} cycle is the boundary of a region that can be tiled by plaquettes. A product of $Z$'s acting on the links of the cycle can be expressed as a product of the enclosed plaquette operators, which acts trivially on the code space. A {\em homologically nontrivial} cycle wraps around the torus and is not the boundary of anything. A product of $Z$'s acting on the links of the cycle preserves the code space, but acts nontrivially on the encoded quantum information. Associated with the two fundamental nontrivial cycles of the torus are encoded operations $\bar Z_1$ and $\bar Z_2$ acting on the two encoded qubits. Similarly, associated with the two dual cycles of the dual lattice are the corresponding encoded operations $\bar X_1$ and $\bar X_2$, as shown in Fig.~\ref{fig:encoded_ops}.

A general error acting on the code block can be expanded in terms of Pauli operators. Therefore, we can characterize the efficacy of error correction by considering how well we can protect the encoded state against Pauli operator errors. With the toric code, $X$ errors (bit flips) and $Z$ errors (phase flips) can be corrected independently; this suffices to protect against general Pauli errors, since a $Y$ error is just a bit flip and a phase flip acting on the same qubit. We may therefore confine our attention to $Z$ errors; the $X$ errors may be dealt with in essentially the same way, but with the lattice replaced by its dual.

\subsection{Perfect measurements and the random-bond Ising model}

To be concrete, suppose that the $Z$ errors are independently and identically distributed, occuring with probability $p$ on each qubit. Noise produces an {\em error chain} $E$, a set of qubits acted upon by $Z$. To diagnose the errors, the code's local check operators are measured at each lattice site, the measurement outcomes providing a ``syndrome'' that we may use to diagnose errors. However, the syndrome is highly ambiguous. It does not completely characterize where the errors occured; rather it only indicates whether the number of damaged qubits adjacent to each site is even or odd. That is, the measurement determines the {\em boundary} $\partial E$ of the error chain $E$.

To recover from the damage, we choose a {\em recovery chain} $E'$ that has the same boundary as the measured boundary of $E$, and apply $Z$ to all the qubits of $E'$. Since $\partial E=\partial E'$, the chain $D=E+E'$ is a cycle with $\partial D=0$. Now, if $D$ is homologically trivial, then the recovery  successfully protects the two encoded qubits --- the effect of the errors together with the recovery step is just to apply a product of check operators, which has trivial action on the code space. But if $D$ is homologically nontrivial, then recovery fails --- the encoded quantum information suffers an error.

Error recovery succeeds, then, if we can guess the homology class of the randomly generated chain $E$, knowing only its boundary $\partial E$ --- we succeed if our guess $E'=E+D$ differs from $E$ by a homologically trivial cycle $D$. If the error rate $p$ is below a certain critical value $p_c$ called the {\em accuracy threshold}, it is possible to guess correctly, with a probability of failure that approaches zero for a sufficiently large linear size $L$ of the lattice. But if $p$ is above $p_c$, the failure probability approaches a nonzero constant as $L\to\infty$. The numerical value of $p_c$ is of considerable interest, since it characterizes how reliably quantum hardware must perform for a quantum memory to be robust.

Let ${\rm prob}(E)$ denote the probability that the error chain is $E$, and let ${\rm prob}[(E+D)|E]$ denote the normalized conditional probability for error chains $E'=E+D$ that have the same boundary as $E$.  Then, the probability of error per qubit lies below threshold if and only if, in the limit $L\to \infty$,
\begin{equation}
\sum_E {\rm prob}(E)\cdot \sum_{D~{\rm nontrivial}} {\rm prob}[(E+D)|E]=0~.
\label{ec_cond}
\end{equation}
Eq.~(\ref{ec_cond}) says that error chains that differ from the actual error chain by a homologically nontrivial cycle have probability zero. Therefore, the outcome of the measurement of the check operators is sure to point to the correct homology class, in the limit of an arbitrarily large code block.

This criterion is identical to the criterion for long-range order in the two-dimensional RBIM, along the Nishimori line. The error chain $E$ can be identified with the chain of antiferromagnetic bonds of a sample, bounded by Ising vortices that are pinned down by the measurement of the local check operators. The ensemble of all the chains $\{E'\}$ with a specified boundary can be interpreted as a thermal ensemble. If the temperature $T$ and the error rate $p$ obey Nishimori's relation, then the chain $E'$ and the chain $E$ have the same bond concentration. At low temperature along the Nishimori line, the cycle $D=E+E'$ contains no large connected loops for typical samples and typical thermal fluctuations --- the spin system is magnetically ordered and error recovery succeeds with high probability. But at higher temperature, the quenched chain $E$ and the thermal chain $E'$ fluctuate more vigorously. At the Nishimori point, $D$ contains loops that ``condense,'' disordering the spins and compromising the effectiveness of error correction. Thus, the critical concentration $p_c$ at the Nishimori point of the two-dimensional RBIM coincides with the accuracy threshold for quantum memory using toric codes (where $p_c$ is the largest acceptable probability for either an $X$ error or a $Z$ error).

The optimal recovery procedure is to choose a recovery chain $E'$ that belongs to the most likely homology class, given the known boundary of the chain $\partial E'=\partial E$. For $p < p_c$, the probability distribution has support on a single class in the limit $L\to \infty$, and the optimal recovery procedure is sure to succeed. In the language of the RBIM, for a given sample with antiferromagnetic chain $E$, a chain $E'$ of excited bonds can be classified according to the homology class to which the cycle $D=E+E'$ belongs, and a free energy can be defined for each homology class.  For $p<p_c$ along the Nishimori line, the trivial homology class has lowest free energy, and the free energy cost of choosing a different class diverges as $L\to \infty$.

An alternative recovery procedure is to choose the single most likely recovery chain $E'$, rather than a chain that belongs to the most likely {\em class}. In the language of the RBIM, this most likely recovery chain $E'$ for a given sample is the set of excited links that minimizes energy rather than free energy. This energy minimization procedure is sure to succeed if the error rate is $p<p_{c0}$, where $p_{c0}$ is the critical bond concentration of the RBIM at $T=0$. Since minimizing energy rather than free energy need not be optimal, we see that $p_{c0} \le p_c$. However, the energy minimization procedure has advantages: it can be carried out efficiently using the Edmonds perfect matching algorithm \cite{edmonds,barahona_match}, and without any prior knowledge of the value of $p$. 

\subsection{Imperfect measurement and the random-plaquette gauge model}
\label{sec:memory_rpgm}

But the RBIM applies only to an unrealistic situation in which the outcomes of measurements of check operators are known with perfect accuracy. Since these are four-qubit measurements, they must be carried out with a quantum computer and are themselves degraded by noise. To obtain reliable information about the positions of the Ising vortices, we must repeat the measurements many times, assembling a measurement history from which we can infer the world lines of the vortices in three-dimensional spacetime. 

\begin{figure}
\begin{center}
\leavevmode
\epsfxsize=3in
\epsfbox{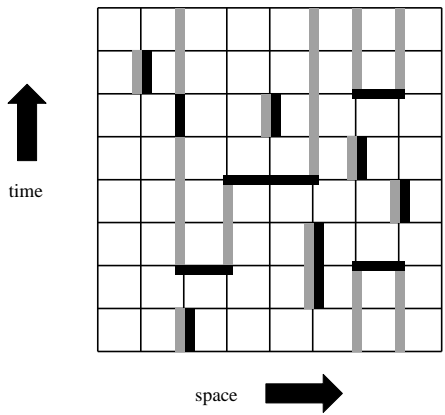}
\end{center}
\caption{An error history shown together with the syndrome history that it generates, for the toric code. For clarity, the three-dimensional history of the two-dimensional code block has been compressed to two dimensions. Qubits reside on plaquettes, and four-qubit check operators are measured at each vertical link. Links where errors have occured are darkly shaded, and links where the syndrome is nontrivial are lightly shaded. Errors on horizontal links indicate where a qubit flipped between successive syndrome measurements, and errors on vertical links indicate where the syndrome measurement was wrong. Vertical links that are shaded both lightly and darkly are locations where a nontrivial syndrome was found erroneously. The chain $S$ of lightly shaded links (the syndrome) and the chain $E$ of darkly shaded links (the errors) both have the same boundary.}
\label{fig:error_history}
\end{figure}

To visualize the world lines in three dimensions, consider a three-dimensional simple cubic lattice on $T^2\times R$, where $T^2$ is the two-torus and $R$ is the real line. The error operation acts at each integer-valued time $t$, and check operators are measured between each $t$ and $t+1$. Qubits in the code block are associated with timelike plaquettes, those lying in the $tx$ and $ty$ planes.  A qubit error that occurs at time $t$ is associated with a horizontal (spacelike) bond that lies in the time slice labeled by $t$.  An error in the measurement of a check operator at site $j$ between time  $t$ and time $t+1$ is associated with the vertical (timelike) bond connecting site $j$ at time $t$ and site $j$ at time $t+1$. Qubit errors on horizontal bonds occur with probability $p$, and measurement errors on vertical links occur with probability $q$. The set of all errors, both horizontal and vertical, defines a one-chain $E$, shown  darkly shaded in Fig.~\ref{fig:error_history}. The set of all syndrome measurements with nontrivial outcomes (those where the observed value of the check operator is $-1$ rather than $+1$) defines a (vertical) one-chain $S$, shown lightly shaded in Fig.~\ref{fig:error_history}.
The chains $E$ and $S$ share the same boundary; therefore the (possibly faulty) measurements of the check operators reveal the boundary of the error chain $E$. 

Error recovery succeeds if we can guess the homology class of the error chain $E$, given knowledge of its boundary $\partial E$; that is, we succeed if our guess $E'=E+D$ differs from $E$ by a cycle $D$ that is homologically trivial on $T^2\times R$. Thus, the accuracy threshold can be mapped to the confinement-Higgs transition of the RPGM. The error one-chain $E$ on the dual lattice becomes the set of wrong-sign plaquettes on the lattice; its boundary points are magnetic monopoles, whose locations are determined by the measurements of local check operators. Since $q$ need not equal $p$, the gauge model can be anisotropic --- on the original lattice, the concentration of spacelike wrong-sign plaquettes is $q$ (spacelike plaquettes are dual to timelike bonds) and the concentration of timelike wrong-sign plaquettes is $p$ (timelike plaquettes are dual to spacelike bonds). The ensemble of error chains $\{E'\}$ that have the same boundary as $E$ becomes the thermal ensemble determined by an anisotropic Hamiltonian, with the coupling $K_{\rm space}$ on spacelike plaquettes obeying the Nishimori relation $K_{\rm space}=K_q$ and the coupling $K_{\rm time}$ on timelike plaquettes the relation $K_{\rm time}=K_p$. 

For small $p$ and $q$, the cycle $D=E+E'$ contains no large connected loops for typical samples and typical thermal fluctuations --- the gauge system is magnetically ordered and error recovery succeeds with high probability. But there is a critical curve in the $(p,q)$ plane where the  magnetic flux tubes ``condense,'' magnetically disordering the system and compromising the effectiveness of error correction. For the sort of error model described in \cite{topo_memory}, the qubit error rate and the measurement error rate are comparable, so the isotropic model with $p=q$ provides useful guidance. For that case, the critical concentration $p_c$ at the Nishimori point of the three-dimensional RPGM coincides with the accuracy threshold for quantum memory using toric codes (where $p_c$ is the largest acceptable probability for an $X$ error, a $Z$ error, or a measurement error). In the extreme anisotropic limit $q\to 0$, flux on spacelike plaquettes is highly suppressed, and the timelike plaquettes on each time slice decouple, with each slice described by the RBIM.

For both the 2D RBIM and the 3D (isotropic) RPGM, we may infer (as Nishimori argued for the RBIM \cite{nish_line}) that the phase boundary lies in the region $p\le p_c$, i.e., does not extend to the right of the Nishimori point. From the perspective of the error recovery procedure, this property reflects that the best hypothesis about the error chain, when its boundary is known, is obtained by sampling the distribution ${\rm prob}[(E+D)|E]$. Thus, for each value of $p$, the fluctuations of $D$ are best controlled (the spins or gauge variables are least disordered) by choosing the temperature on the Nishimori line. For $p>p_c$ the magnetization of the 2D RBIM vanishes on the Nishimori line, and so must vanish for all $T$. A similar remark applies to the Wilson-loop order parameter of the 3D RPGM.

In particular, the critical value of $p$ on the $T=0$ axis (denoted $p_{c0}$) provides a lower bound on $p_c$. Rigorous arguments in \cite{topo_memory} established that $p_{c0}\ge .0373$ in the 2D RBIM and $p_{c0}\ge .0114$ in the 3D RPGM. (A similar lower bound for the 2D RBIM was derived by Horiguchi and Morita many years ago \cite{horiguchi}.) We have estimated the value of $p_{c0}$ using numerical simulations that we will now describe.

\section{Numerics}

\subsection{Method}
For the RBIM in two dimensions (but not in higher dimensions), and for the RPGM in three dimensions (but not in higher dimensions), it is numerically tractable to study the phase transition on the $T=0$ axis. Specifically, for the RBIM, we proceed as follows: Consider an $L\times L$ lattice on the torus, and generate a sample by choosing a random $\tau_{ij}$ at each bond (where $\tau_{ij}=-1$ occurs with probability $p$). Consider, for this sample, the one-chain $E$ on the dual lattice containing bonds with $\tau_{ij}=-1$, and compute its boundary $\partial E$ to locate the Ising vortices. 

Then, to find the ground state of the Hamiltonian for this sample, construct the one-chain $E'$ of the dual lattice, bounded by the Ising vortices, with the minimal number of bonds. This minimization can be carried out in a time polynomial in $L$ using the Edmonds perfect matching algorithm \cite{edmonds,barahona_match}.  (If the ground state is not unique, choose a ground state at random.) Now examine the one-cycle $D=E+E'$ on the torus and compute whether its homology class is trivial. If so, we declare the sample a ``success;'' otherwise the sample is a ``failure.'' Repeat for many randomly generated samples, to estimate the probability of failure $P_{\rm fail}(p)$.

We expect $P_{\rm fail}(p)$ to be discontinuous at $p=p_{c0}$ in the infinite volume limit. For $p<p_{c0}$, large loops in $D$ are heavily suppressed, so that $P_{\rm fail}$ falls exponentially to zero for $L$ sufficiently large compared to the correlation length $\xi$. But for $p> p_{c0}$, arbitrarily large loops are not suppressed, so we anticipate that the homology class is random. Since there are four possible classes, we expect $P_{\rm fail}$ to approach $3/4$ as $L\to\infty$.

This expectation suggests a finite-size scaling ansatz for the failure probability. Let the critical exponent $\nu_0$ characterize the divergence of the correlation length $\xi$ at the critical point $p=p_{c0}$:
\begin{equation}
\xi\sim |p-p_{c0}|^{-\nu_0}~.
\end{equation}
For a sufficiently large linear size $L$ of the sample, the failure probability should be controlled by the ratio $L/\xi$; that is, it is a function of the scaling variable 
\begin{equation}
x= (p-p_{c0})L^{1/\nu_0}~.
\end{equation}
Thus the appropriate ansatz is
\begin{equation}
\label{ansatz}
P_{\rm fail}\sim {3\over 4} f(x)~,
\end{equation}
where the function $f$ has the properties
\begin{equation}
\lim_{x\to -\infty} f(x)=0~,\quad \lim_{x\to\infty} f(x) =1~.
\end{equation}
Though the scaling ansatz should apply asymptotically in the limit of large $L$, there are systematic corrections for finite $L$ that are not easily estimated. 

According to eq.~(\ref{ansatz}), the failure probability at $p=p_{c0}$ has a universal value $(3/4) f(0)$ that does not depend on $L$. Thus, by plotting $P_{\rm fail}$ vs. $p$ for various values of $L$, we can estimate $p_{c0}$ by identifying the value of $p$ where all the curves cross. To find $\nu_0$, we observe that
\begin{equation}
\log\left({\partial P_{\rm fail}\over\partial p}\Big|_{p=p_{c0}}\right)= {1\over \nu_0}\log L +{\rm constant}~.
\end{equation}
Hence, if we estimate the slope of $P_{\rm fail}$ at $p=p_{c0}$, we can extract  $\nu_0$ from a linear fit to a plot of $\log({\rm slope})$ vs. $\log L$.

The three-dimensional RPGM can be analyzed by the same method. A sample is generated by randomly choosing $\tau_P$ on each plaquette of an $L^3$ cubic lattice on the 3-torus. The wrong-sign plaquettes define a one-chain $E$ on the dual lattice, whose boundary defines the locations of the magnetic monopoles. The ground state of the sample is constructed by finding the one-chain $E'$ with the same boundary that has the minimal length, and the one-cycle $D=E+E'$ is examined to determine if it is homologically trivial. Since there are eight homology classes on the 3-torus, the scaling ansatz becomes
\begin{equation}
\label{ansatz_gauge}
P_{\rm fail}\sim {7\over 8} \tilde f(x)~,
\end{equation}
and $p_{c0}$ and $\nu_0$ are estimated as described above.

For the RBIM in three dimensions, or the RPGM in four dimensions, $E$ and $E'$ become two-chains. To construct the ground state, then, we must find the minimal two-dimensional surface that has a specified boundary. Unfortunately, this problem is known to be NP-hard \cite{barahona} and so appears to be computationally intractable.

Detailed numerical studies of the two-dimensional RBIM in the vicinity of the Nishimori point have been done earlier by other authors \cite{honecker,chalker}, using methods that are not very effective at low temperature. The $T=0$ phase transition has been studied using methods related to ours \cite{barahona_match,kawashima}, but with less numerical accuracy. As far as we know, numerical studies of the RPGM have not been previously attempted.

\subsection{Random-bond Ising model}

\begin{figure}
\begin{center}
\leavevmode
\epsfxsize=3in
\epsfbox{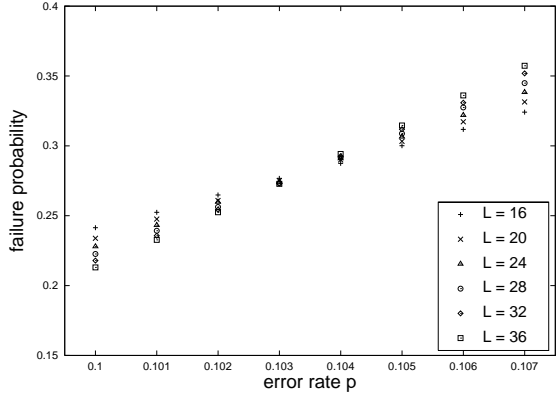}
\end{center}
\caption{The failure probability $P_{\rm fail}$ as a function of the error probability $p$ for linear size $L=16,20,24,28,32,36$, in the two-dimensional random-bond Ising model. Each data point was generated by averaging $10^6$ samples.}
\label{fig:2D_data}
\end{figure}

\begin{figure}
\begin{center}
\leavevmode
\epsfxsize=3in
\epsfbox{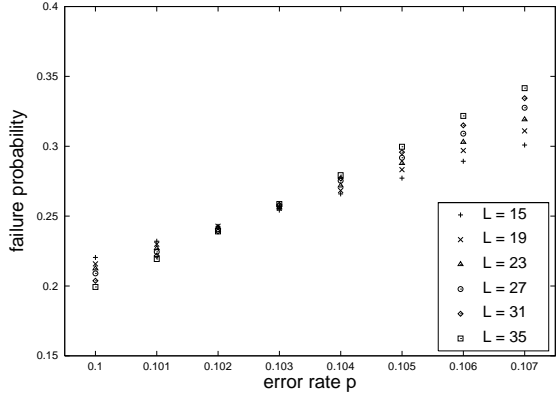}
\end{center}
\caption{The failure probability $P_{\rm fail}$ as a function of the error probability $p$ for linear size $L=15,19,23,27, 31,35$, in the two-dimensional random-bond Ising model. Each data point was generated by averaging $10^6$ samples.}
\label{fig:2D_data_odd}
\end{figure}

We measured $P_{\rm fail}$ by generating $10^6$ samples for each value of $L$ from 2 to 36, and for each value of $p$ increasing in increments of .001 from .100 to .107; in addition we generated $10^6$ samples at $L=37,38,40,42$ for $p=.102,.103,.104$. Values of $P_{\rm fail}$ for even $L$ lie slightly but systematically above the values for odd $L$ at the same $p$; therefore we analyzed the data for even and odd $L$ separately. Data for $L=16,20,24,28,32,36$ are shown in Fig.~\ref{fig:2D_data}, and data for $L=15,19,23,27,31,35$ are shown in Fig.~\ref{fig:2D_data_odd}. Crudely, the point of concordance of the data sets provides an estimate of $p_{c0}$, while the trend of the data with $L$ determines the exponent $\nu_0$.

We did a global fit of the data to the form
\begin{equation}
P_{\rm fail}= A + Bx + Cx^2~, 
\end{equation}
where $x= (p-p_{c0})L^{1/\nu_0}$, adopting a quadratic approximation to the scaling function $f(x)$ in the vicinity of $x=0$. (In the range of $x$ we considered, the quadratic term is small but not quite negligible.) For even $L$ ranging from 22 to 42, our fit found
\begin{eqnarray}
&p_{c0}= &.10330 \pm .00002~, \nonumber\\
&\nu_0=&1.49 \pm .02~,
\end{eqnarray}
where the quoted errors are one-sigma statistical errors. For odd $L$ ranging from 21 to 37, our fit found
\begin{eqnarray}
&p_{c0}= &.10261 \pm .00003~, \nonumber\\
&\nu_0=&1.46 \pm .02~.
\end{eqnarray}
The discrepancy between the values of $p_{c0}$ for even and odd $L$ indicates a nonnegligible finite-size effect.

\begin{figure}
\begin{center}
\leavevmode
\epsfxsize=3in
\epsfbox{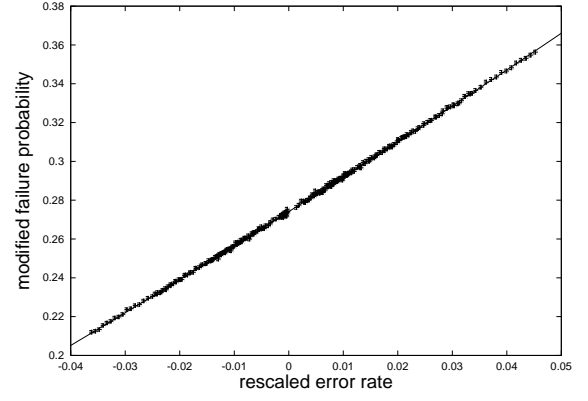}
\end{center}
\caption{The failure probability $P_{\rm fail}$, with the nonuniversal correction of Eq.~(\ref{ansatz_with_correction}) subtracted away, as a function of the scaling variable $x=(p-p_{c0})L^{1/\nu_0}$ for the two-dimensional random-bond Ising model, where $p_{c0}$ and $\nu_0$ are determined by the best fit to the data. A two-sigma error bar is shown for each point. The data for values of $L$ from 2 to 42 lie on a single line, indicating that the (small) scaling violations are well accounted for by our ansatz.}
\label{fig:2D_rescaled}
\end{figure}

On closer examination, we see evidence for small but detectable violations of our scaling ansatz in both the even and odd data sets. These violations are very well accounted for by the modified ansatz
\begin{eqnarray}
\label{ansatz_with_correction}
P_{\rm fail}&=& A + Bx + Cx^2 \nonumber\\
&&+ \cases{D_{\rm even}\cdot L^{-1/\mu_{\rm even}}~& ($L$ {\rm ~even})~,\cr D_{\rm odd}\cdot L^{-1/\mu_{\rm odd}}~& ($L$ {\rm ~odd})~,\cr} 
\end{eqnarray}
which includes a nonuniversal additive correction to $P_{\rm fail}$ at criticality, different for even and odd sizes. Fitting the modified ansatz to the data for even $L$ ranging from 2 to 42, we find
\begin{eqnarray}
\label{even_fit_with_D}
&p_{c0}= &.10309 \pm .00003~, \nonumber\\
&\nu_0=&1.461 \pm .008~,\nonumber~\\
&D_{\rm even}=&0.165\pm .002~,\quad \mu_{\rm even}=0.71\pm .01~.
\end{eqnarray}
Fitting to the data for odd $L$ ranging from 3 to 37, we find
\begin{eqnarray}
\label{odd_fit_with_D}
&p_{c0}= &.10306 \pm .00008~, \nonumber\\
&\nu_0=&1.463 \pm .006~,\nonumber~\\
&D_{\rm odd}=&-.053 \pm .003~,\quad \mu_{\rm odd}=2.1\pm .3~.
\end{eqnarray}
In Fig.~(\ref{fig:2D_rescaled}) we show the data for all values of $L$ and $p$; using the values of $p_{c0}$, $\nu_0$, $D$, and $\mu$ found in our fits, we have plotted $P_{\rm fail}$, with the nonuniversal correction of Eq.~(\ref{ansatz_with_correction}) subtracted away, as a function of the scaling variable $x= (p-p_{c0})L^{1/\nu_0}$. All of the data lie on a single line, indicating that residual scaling violations are quite small. Furthermore, the agreement between the values of $p_{c0}$ and $\nu_0$ extracted from the even and odd data sets, which were fit independently, indicates that our extrapolation to large $L$ is reasonable, and that the statistical errors in Eq.~(\ref{even_fit_with_D},\ref{odd_fit_with_D}) do not seriously underestimate the actual errors in our measurement. A plausible conclusion is that
\begin{eqnarray}
&p_{c0}= &.1031 \pm .0001~, \nonumber\\
&\nu_0=&1.46 \pm .01~.
\end{eqnarray}

An earlier measurement reported by Kawashima and Rieger found \cite{kawashima}
\begin{eqnarray}
&p_{c0}= &.104 \pm .001~,\nonumber\\
&\nu_0=&1.30 \pm .02 ~;
\end{eqnarray}
their value of $p_{c0}$, but not of $\nu_0$, is compatible with ours. An important reason why our value of $p_{c0}$ has a smaller statistical error than theirs is that they computed a different observable (the domain wall energy) for which the finite-size scaling analysis is more delicate than for the failure probability (another critical scaling exponent is involved).

In a recent study of the Nishimori point, Merz and Chalker found \cite{chalker}
\begin{eqnarray}
&p_c= &.1093 \pm .0002~,\nonumber\\
&\nu=&1.50 \pm .03 ~.
\end{eqnarray}
There is a clear discrepancy between the values of $p_c$ and $p_{c0}$, in disagreement with the conjecture of Nishimori \cite{nish_vert} and Kitatani \cite{kitatani}. Evidence for a reentrant phase diagram has also been found by Nobre \cite{nobre}, who reported
\begin{equation}
p_{c0}= .1049 \pm .0003~.
\end{equation}

In principle, the phase transitions at $T=0$ and at the Nishimori point could be in different universality classes, so that the critical exponents $\nu_0$ and $\nu$ could have different values. However, our measurement of $\nu_0$ at $T=0$ is consistent with the value of $\nu$ at the Nishimori point reported by Merz and Chalker \cite{chalker}.

\subsection{Random-plaquette gauge model}

\begin{figure}
\begin{center}
\leavevmode
\epsfxsize=3in
\epsfbox{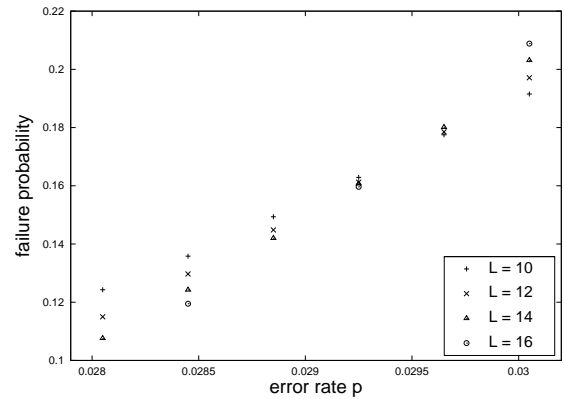}
\end{center}
\caption{The failure probability $P_{\rm fail}$ as a function of the error probability $p$ for linear size $L=10,12,14,16$, in the three-dimensional random-plaquette gauge model. Each data point was generated by averaging $10^6$ samples.}
\label{fig:3D_data}
\end{figure}

We measured $P_{\rm fail}$ by generating $10^6$ samples for each value of $L$ from 9 to 14, and for each value of $p$ increasing in increments of .0004 from .02805 to .03005; in addition we generated $10^6$ samples at $L=15,16$ for $p=.02845,.02925,.03005$. Values of $P_{\rm fail}$ for even $L$ lie slightly but systematically above the values for odd $L$ at the same $p$; therefore we analyzed the data for even and odd $L$ separately. Data for even $L$ are shown in Fig.~\ref{fig:3D_data}. Crudely, the point of concordance of the data sets provides an estimate of $p_{c0}$, while the trend of the data with $L$ determines the exponent $\nu_0$.

We did a global fit of the data to the form
\begin{equation}
P_{\rm fail}= A + Bx +Cx^2~, 
\end{equation}
where $x=(p-p_{c0})L^{1/\nu_0}$, adopting a quadratic approximation to the scaling function $f(x)$ in the vicinity of $x=0$. For $L$ ranging from 9 to 16, our fit found
\begin{eqnarray}
&p_{c0}= .02937 \pm .00002~,~ &\nu_0=0.974 \pm .026~(L ~{\rm even})~,\nonumber\\
&p_{c0}= .02900 \pm .00001~,~ &\nu_0=1.025 \pm .016~(L ~{\rm odd})~,
\end{eqnarray}
where the quoted errors are one-sigma statistical errors. The results for even and odd $L$ are incompatible, indicating a nonnegligible finite-size effect.

\begin{figure}
\begin{center}
\leavevmode
\epsfxsize=3in
\epsfbox{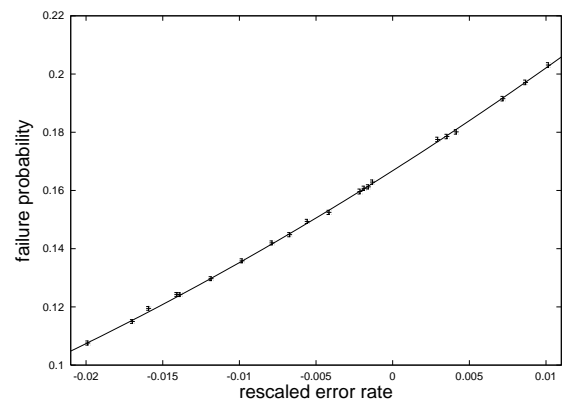}
\end{center}
\caption{The failure probability $P_{\rm fail}$ as a function of the scaling variable $x=(p-p_{c0})L^{1/\nu_0}$ for the random-plaquette gauge model, where $p_{c0}$ and $\nu_0$ are determined by the best fit to the data. A two-sigma error bar is shown for each point. The data for all even values of $L$ from 10 to 16 lie on a single curve, indicating that scaling violations are small.}
\label{fig:3D_rescaled}
\end{figure}

We believe that our analysis for even $L$ is likely to be more reliable; finite size effects are enhanced for odd $L$, the case in which the failure probability is smaller. All of the even-$L$ data are shown in Fig.~\ref{fig:3D_rescaled}, with $P_{\rm fail}$ plotted as a function of  $x=(p-p_{c0})L^{1/\nu_0}$, where $p_{c0}$ and $\nu_0$ are determined by our fit. The data fit a single curve, indicating that scaling violations are small. (Scaling violations are more discernable in the odd-$L$ data set.) A reasonable conclusion is that
\begin{eqnarray}
&p_{c0}= &.0293 \pm .0002~,\nonumber\\
&\nu_0=&1.00 \pm .05 ~.
\end{eqnarray}

\subsection{The failure probability at finite temperature}
Our numerical studies of the RBIM and the RPGM were restricted to the $T=0$ axis. We calculated the failure probability to estimate the critical disorder strength $p_{c0}$ and the critical exponent $\nu_0$. Here we will describe how the calculation of the failure probability could be extended to nonzero temperature.

To calculate the failure probability in the zero-temperature RBIM, we generate a sample by specifying a one-chain $E$ of antiferromagnetic links, and then we construct the one-chain $E'$ of minimal energy with the same boundary as $E$. Failure occurs if the cycle $D=E+E'$ is homologically nontrivial.

At nonzero temperature we should construct $E'$ to belong to the homology class that minimizes free energy rather than energy. For a given sample with antiferromagnetic one-chain $E$, the free energy $F(E,h)$ of homology class $h$ is found by summing over domain wall one-chains $\{E'\}$ such that $E+E'\in h$:
\begin{equation}
\exp[-\beta F(E,h)]= Z(E,h) = \sum_{E':E+E'\in h}e^{-\beta H_E}~,
\end{equation}
where $H_E$ denotes the Hamiltonian eq.~(\ref{Ising_Hamiltonian}) with antiferromagnetic chain $E$.
If the trivial homology class $h=e$ has the lowest free energy, then the sample is a ``success;'' otherwise it is a ``failure.'' We can estimate the failure probability $P_{\rm fail}(p,T)$ by randomly generating many samples, and determining for each whether it is a success or a failure.

For the random bond Ising model on a torus, the sum eq.~(\ref{partition_function}) includes only the chains $E'$ such that $E+E'$ is in the trivial homology class. To sum over the class $h$, we can augment $E$ by adding to it a representative of $h$. For each $h$, we can compute
\begin{equation}
{Z(E,h)\over Z(E,e)}=\exp\big[-\beta \big(F(E,h) -F(E,e)\big)\big]~;
\end{equation}
the sample $E$ is a success if this ratio of partition functions is less than one for each $h\ne e$.

The ratio is the thermal expectation value $\langle {\cal O}_h\rangle_{K}$ of an observable ${\cal O}_h$ that ``inserts a domain wall'' wrapping around a cycle $C$ representing $h$. That is, the effect of ${\cal O}_h$ is to flip the sign of the bond variable $\tau_{ij}$ for each bond $\langle ij\rangle$ in $C$: 
\begin{equation}
{\cal O}_h=\exp\left[-2K\sum_{\langle ij \rangle \in C}\tau_{ij}S_iS_j\right]~.
\end{equation}
In principle, we could measure $\langle {\cal O}_h\rangle_{K}$ by the Monte Carlo method, generating typical configurations in the thermal ensemble of $H_E$, and evaluating ${\cal O}_h$ in these configurations. 
Unfortunately, this method might not produce an accurate measurement, because the configurations that dominate  $\langle {\cal O}_h\rangle_{K}$ may be exponentially rare in the thermal ensemble --- a configuration with excited bonds on $C$ can have an exponentially large value of ${\cal O}_h$ that overcomes exponential Boltzmann suppression.

One solution to this problem is to express $Z(E,h)/Z(E,e)$ as a product of quantities, each of which {\em can} be evaluated accurately by Monte Carlo. Let $\{e=P_0,P_1, P_2, \dots P_{k-1},P_k=C\}$ be a sequence of open chains interpolating between the empty chain and the cycle $C$, where $P_{j+1}-P_j$ contains just a single bond. We may write
\begin{equation}
{Z(E,h)\over Z(E,e)}={Z(E,P_1)\over Z(E,P_0)}\cdot {Z(E,P_2)\over Z(E,P_1)}\cdot \cdots \cdot {Z(E,P_k)\over Z(E,P_{k-1})}~.
\end{equation}
Each ratio ${Z(E,P_{j+1})/ Z(E,P_j)}$ is the expectation value of an operator that acts on a single bond, evaluated in the thermal ensemble of the Hamiltonian with antiferromagnetic bonds on the chain $E+P_j$;  this expectation value can be evaluated by Monte Carlo with reasonable computational resources. (For an application of this trick in a related setting, see \cite{forcrand}.)

Using this method, we can determine whether ${Z(E,h)/ Z(E,e)}$ exceeds one for any $h\ne e$ and hence whether the sample $E$ is a success or a failure. Generating many samples, we can estimate $P_{\rm fail}(p,T)$. In principle, then we can calculate the failure probability for the optimal recovery scheme, in which $p$ and $T$ obey Nishimori's relation. By a similar method, we can calculate the failure probability for the RPGM. However, we have not attempted this calculation.

\section{Conclusions}

The three-dimensional random-plaquette gauge model, and the analogous antisymmetric tensor models in higher dimensions, provide new examples of multicritical points with strong disorder. These models have phase diagrams that qualitatively resemble the phase diagram of the two-dimensional random-bond Ising model. 

Our results indicate that the boundary between the ferromagnetic and paramagnetic phases of the RBIM is reentrant rather than vertical below the Nishimori line. If the disorder strength $p$ satisfies $p_{c0} < p < p_c$, then the ground state of the spin system does not have long-range order. As the temperature $T$ increases with $p$ fixed, long-range order is first restored, and then lost again as the temperature increases further. At $T=0$ the spins are frozen in a disordered state driven by quenched randomness. But apparently this ground state is entropically unfavorable --- at low but nonzero temperature typical states in the thermal ensemble have long-range ferromagnetic order.

This behavior seems less remarkable when considered from the viewpoint of our error recovery protocol. For given $p$ and a specified error syndrome, the recovery method with optimal success probability proceeds by inferring the most likely homology class of errors consistent with the syndrome. There is no {\em a priori} reason for the most likely single error pattern (the ground state) to belong to the most likely error homology class (the class with minimal free energy) even in the limit of a large sample. Our numerical results indicate that for error probability $p$ such that $p_{c0} < p <p_c$, the probability that the ground state does not lie in the most likely homology class remains bounded away from zero as $L\to\infty$.

In our numerical studies of the RBIM and RPGM at zero temperature, we have computed a homological observable, the failure probability. This observable has advantages over, say, the domain wall energy, because it obeys a particularly simple finite-size-scaling ansatz. Therefore, we have been able to determine the critical disorder strength $p_{c0}$  and the critical exponent $\nu_0$ to good accuracy with relatively modest computational resources. 

Not surprisingly, our numerical values for $p_{c0}$ are notably larger than rigorous lower bounds derived using crude combinatoric arguments in \cite{topo_memory}: $p_{c0} \approx .1031$ compared with the bound $p_{c0} \ge .0373$ in the RBIM, and $p_{c0}\approx .0293$ compared with $p_{c0} \ge .0114$ in the RPGM.

The zero-temperature critical disorder strength $p_{c0}$ is a lower bound on the value of the critical disorder strength $p_c$ along the Nishimori line, and of special interest because of its connection with the accuracy threshold for robust storage of quantum information. Our result means that stored quantum data can be preserved with arbitrarily good fidelity if, in each round of syndrome measurement, qubit errors and syndrome measurement errors are independently and identically distributed, with error probability per qubit and per syndrome bit both below $2.9\%$. For qubit errors and measurement errors occuring at differing rates, an accuracy threshold could be inferred by analyzing an anisotropic random-plaquette gauge model, with differing disorder strength for horizontal and vertical plaquettes. Relating these threshold error rates to fidelity requirements for quantum gates requires further analysis of the sort discussed in \cite{topo_memory}.

We have also measured the critical exponent $\nu_0$ that controls the divergence of the correlation length as $p$ approaches $p_{c0}$, finding $\nu_0\approx 1.46$ in the RBIM and $\nu_0 \approx 1.0$ in the RPGM. The value of $\nu_0$ is also relevant to the efficacy of quantum error correction --- through its connection with finite-size scaling, $\nu_0$ determines how large the code block should be to achieve a specified storage fidelity, for $p$ less than but close to $p_{c0}$.

Quantum computers are believed to be more powerful than classical computers --- classical computers are unable to simulate quantum computers efficiently. The accuracy threshold for quantum memory is a fascinating phase transition, separating a low-noise quantum phase from a high-noise classical phase. In this paper, we have described one way to analyze this phase transition using methods from traditional statistical physics. Furthermore, the connection with quantum memory provides an enlightening new perspective on local spin and gauge systems with strong quenched disorder.

\acknowledgments

We gratefully acknowledge helpful discussions and correspondence with John Chalker, Tom Gottschalk, Alexei Kitaev, Hidetsugu Kitatani, Andreas Ludwig, Paul McFadden, Hidetoshi Nishimori, and Frank Porter. We particularly thank Andrew Landahl, Nathan Wozny, and Zhaosheng Bao for valuable advice and assistance. This work has been supported in part by the Department of Energy under Grant No. DE-FG03-92-ER40701, by the National Science Foundation under Grant No. EIA-0086038, by the Caltech MURI Center for Quantum Networks under ARO Grant No. DAAD19-00-1-0374, and by Caltech's Summer Undergraduate Research Fellowship (SURF) program. 

\renewcommand{\baselinestretch}{0.96}

\end{multicols}

\end{document}